\documentclass[conference]{IEEEtran}
\IEEEoverridecommandlockouts

\usepackage{cite}
\usepackage{titlesec}
\titlespacing*{\section}{0pt}{1.0ex}{0.5ex}
\titlespacing*{\subsection}{0pt}{0.8ex}{0.4ex}
\usepackage[font=small, skip=2pt]{caption}
\usepackage{setspace}
\usepackage{etoolbox}
\AtBeginEnvironment{thebibliography}{\small}

\usepackage{amsmath,amssymb,amsfonts}
\usepackage{dsfont}
\usepackage{graphicx}
\usepackage{textcomp}
\usepackage{xcolor}
\usepackage{cleveref}
\usepackage{amsthm}
\usepackage{bbm}

\usepackage{wrapfig}
\usepackage{booktabs}
\usepackage{multicol}
\usepackage{multirow}
\usepackage{tabularx}
\usepackage{diagbox}

\usepackage{subfigure} 

\usepackage{array}
\usepackage{fancyhdr}

\usepackage{float}

\usepackage{longtable}
\usepackage{tikz}
\usepackage{verbatim}
\usepackage{listings}

\definecolor{mygreen}{RGB}{28,172,0} 
\definecolor{mylilas}{RGB}{170,55,241}

\usepackage{matlab-prettifier}
\usepackage{color}

\usepackage[shortlabels]{enumitem}

\usepackage{algorithm}
\usepackage[noend]{algpseudocode}

\makeatletter
\def\BState{\State\hskip-\ALG@thistlm}
\makeatother

\newtheorem{definition}{Definition}

\def\BibTeX{{\rm B\kern-.05em{\sc i\kern-.025em b}\kern-.08em
    T\kern-.1667em\lower.7ex\hbox{E}\kern-.125emX}}
\begin{document}

\title{\bf \LARGE  ADAPT: A Game-Theoretic and Neuro-Symbolic Framework for Automated Distributed Adaptive Penetration Testing
\thanks{\noindent \IEEEauthorrefmark{2}Authors contributed equally, \IEEEauthorrefmark{1}Corresponding author, \textit{Authors belong to Department of Electrical and Computer Engineering} \textit{New York University, New York 11201, USA}\{hl4155, yg2047, qz494\}@nyu.edu}
}\vspace{-6mm}



\author{
\IEEEauthorblockN{Haozhe Lei\IEEEauthorrefmark{1}\IEEEauthorrefmark{2}, Yunfei Ge \IEEEauthorrefmark{2},
Quanyan Zhu}
\vspace{-10mm}
}

\maketitle

\begin{abstract}
The integration of AI into modern critical infrastructure systems, such as healthcare, has introduced new vulnerabilities that can significantly impact workflow, efficiency, and safety. Additionally, the increased connectivity has made traditional human-driven penetration testing insufficient for assessing risks and developing remediation strategies. Consequently, there is a pressing need for a distributed, adaptive, and efficient automated penetration testing framework that not only identifies vulnerabilities but also provides countermeasures to enhance security posture. This work presents ADAPT, a game-theoretic and neuro-symbolic framework for automated distributed adaptive penetration testing, specifically designed to address the unique cybersecurity challenges of AI-enabled healthcare infrastructure networks. We use a healthcare system case study to illustrate the methodologies within ADAPT. The proposed solution enables a learning-based risk assessment. Numerical experiments are used to demonstrate effective countermeasures against various tactical techniques employed by adversarial AI.
\end{abstract}


\section{Introduction}
Modern artificial intelligence (AI), such as machine learning (ML) technologies, are becoming increasingly integrated into many infrastructures, including smart transportation systems and healthcare infrastructures. In healthcare, they have shown the potential to help healthcare infrastructure in patient scheduling \cite{scheduling_ML, scheduling_ML2}, pathological analysis \cite{Ahsan2022MachineLearningDiagnosis}, and care management \cite{medicalmerl}. While there are significant benefits, there are concerns regarding zero-day vulnerabilities and the expanded attack surface. 


Penetration testing is a valuable ethical hacking method for uncovering vulnerabilities in increasingly complex infrastructures and devising remediation strategies. As these infrastructures become more complex, with millions of interconnected devices, scalability emerges as a critical challenge. It is essential to develop a distributed, modular, and automated approach that addresses device-level testing needs while considering global influences through interconnectivity. Another challenge stems from the dynamic nature of networked devices and their vulnerabilities. There is a growing need for adaptive and automated approaches to continuously update the vulnerability landscape, ensuring that threats are exhaustively identified, risks accurately assessed, and remediation measures properly applied. The third challenge arises from the integration of AI capabilities into the infrastructure. The emergence of adversarial AI/ML introduces new and evolving threat vectors, which are designed to evade detection and testing. There is a need for the development of automated and strategic approaches that can intelligently outmaneuver their evolving nature through continuous knowledge acquisition and learning.
 To this end, we establish a game-theoretic and neuro-symbolic framework for automated distributed adaptive penetration testing (ADAPT).

\begin{figure}[h]
 \vspace{-2mm}   \centering   \includegraphics[width=0.75\linewidth]{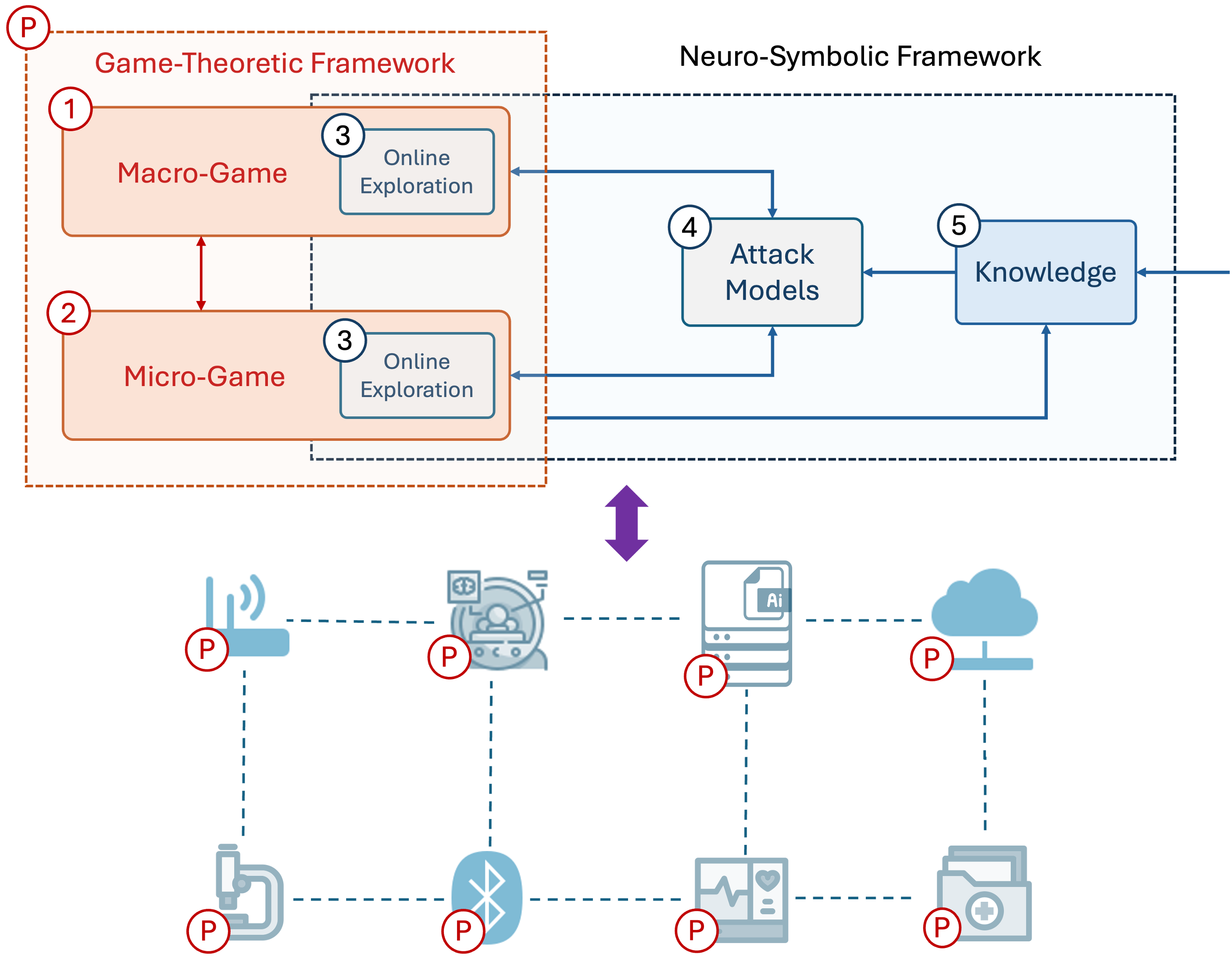}
    \caption{The framework of the ADAPT: The upper half illustrates an online automated adaptation of penetration testing. It integrates game-theoretic and neuro-symbolic frameworks, consisting of five distinct building blocks (to be introduced in Section III). The lower half depicts an example of AI-enabled healthcare infrastructure. This AI-enabled infrastructure presents an expanded attack surface due to interconnectivity and zero-day vulnerabilities. 
    }
    \label{fig:ADAPT}\vspace{-4mm}
\end{figure}

To address the challenges in penetration testing within AI-enabled healthcare infrastructure networks, ADAPT consolidates the game-theoretic framework and the neuro-symbolic framework, shown in the \Cref{fig:ADAPT}. The game-theoretic and neuro-symbolic framework consists of five building blocks. The macro-game and micro-game blocks serve as representations of the given system. Game-theoretic strategies are updated based on the selected attack models through neural learning in the online learning blocks. Different attack models are represented using game trees, which encode relevant attack and defense actions selected from the knowledge. This knowledge contains vulnerabilities of the health infrastructure that are shared among multiple stakeholders in the medical system. When new paths or vulnerabilities are discovered through exploration (e.g., automated fuzzing techniques \cite{Gorbunov2010AutoFuzzAN}) and penetration testing, the knowledge base is generated and updated accordingly. ADAPT helps medical systems evaluate their AI-enabled network for scalability, the impact of reachability, and the exhaustiveness of risk identification, and protects the confidentiality, integrity, and availability of the AI model. The purpose of it is to ensure preparedness against continuously evolving malicious threats such as ransomware and zero-day attacks on healthcare infrastructures. 


\section{Related Works}

Traditional manual penetration testing performed by skilled IT professionals is time-consuming, resource-intensive, and prone to human error \cite{stefinko2016manual}. Current automated penetration testing methods, despite advancements in efficiency, are becoming increasingly non-standard, complex, and resource-intensive. Reinforcement learning (RL) or Markov Decision Process (MDP) based methods \cite{ghanem2019reinforcement,hu2020automated} suffer from the curse of dimensionality, as they define the state space as the collection of all known information for each machine on the network. Partially Observable Markov Decision Process (POMDP) methods \cite{shmaryahu2017partially} face scalability issues, making it unfeasible to model and solve for large networks. 


As state-of-the-art technology, AI models present relatively higher risks compared to traditional methods \cite{Mueller2023UnderstandingRisk}. Several studies have demonstrated that AI models can be easily subjected to malicious data-poisoning attacks in not only regression tasks \cite{YERLIKAYA2022118101} but also image classification \cite{wiyatno2019adversarialexamplesmodernmachine} and robotics control \cite{sample_attack}. In addition to directly degrading the ML model, evidence also points to various other types of attacks, such as DDoS attacks \cite{Head2024DDoSHealthcare} and bypassing attacks \cite{Cylance2019IKillYou}. These findings underscore the significant threat posed by attackers and a wide range of vulnerabilities for the AI-enabled system. 




\section{ADAPT Framework and Methodologies}\label{sec:meth}
This section, we present the ADAPT framework introduced in Figure 1. It consists of a meta-game framework and a neuro-symbolic framework. 
\subsection{Meta-game-based Automated Penetration Testing}

We propose a meta-security game over a network graph where the macro strategic game represents strategic attack activities between nodes, while the micro tactic game details tactic-level attack procedures on each local node. Let the directed graph \(\mathcal{G} = \langle \mathcal{V}, \mathcal{E} \rangle\) represent the target network topology, where \(\mathcal{V}\) is a set of nodes (e.g., server, database, device), and \(\mathcal{E} \subseteq \mathcal{V} \times \mathcal{V}\) is a set of directed edges representing connections (e.g., SSH, RDP, cloud services) from node \(u\) to node \(v\). Self-loops are allowed as they indicate continued exploration of the same node. Let \(v^0 \in \mathcal{V}\) be the initial foothold in the system. As an ethical attacker, the penetration tester aims to explore available information, exploit discovered vulnerabilities, and influence critical assets inside the healthcare infrastructure network.

\textbf{Micro Tactic Games:} To model the interactions between the attacker and the defender at each local node, we use extensive-form game tree to explicitly and visually represent the sequential moves, possible outcomes, and information available at each decision point. We assume that all players have perfect recall; i.e., the player remembers every piece of information that he knows from the past, including his moves, the other player’s moves, or chance moves. The structure of the game tree is inspired by the intrusion kill chain \cite{Hutchins2010IntelligenceDrivenCN}, a concept that outlines the structure of intrusions. This model guides analysis to inform actionable security intelligence.

For each node $v \in \mathcal{V}$ in the network, the Micro Tactic Game (MTG) is defined as $\Gamma^v = \langle \mathcal{N} \cup \{c\}, \mathcal{H}^v, P, \{\mathcal{A}^v_i\}_{i \in \mathcal{N} \cup c}, \sigma^v_c, \{u^v_i\}_{i \in \mathcal{N}}, \mathcal{Z}^v \rangle$. Here, $\mathcal{N} = \{a, d\}$ represents the players, the attacker (a) and the defender (d), while $c$ represents system randomness with fixed policy $\sigma^v_c$. Each vertex $h \in \mathcal{H}^v$ in the game tree represents a sequence of actions, referred to as history. The function $P(h)$ determines whose turn it is at each decision point (attacker, defender, or nature) for a given history vertex $h$. $\mathcal{A}^v_i$ is a set of actions available to player $i$, and $A(h)$ describes the feasible actions for the player at vertex $h$. $\mathcal{Z}^v$ represents the possible outcomes for each tactic in the game, corresponding to the results at the leaf vertices of the game tree. We assume the outcomes are either remaining in the current node or leading to another connected node, denoted as $\mathcal{Z}^v = \{u \mid u \in \mathcal{V}, (v, u) \in \mathcal{E}\}$. Finally, the utility function $u^v_i$ determines the payoff or cost for player $i$ when reaching a certain outcome.

By solving the extensive-form game, we are able to obtain the optimal local penetration for the attacker under possible defense plans. Given the nature's fixed policy (if any) and the plan profile of the attacker and the defender, i.e., $\Phi^v = (\sigma^v_a,\sigma^v_d,\sigma^v_c)$, we denote $\tau^v(z\mid \Phi^v) \in [0,1]$ as the tactic outcome probability, which is the joint probability of reaching that outcome under $\Phi^v$.

\textbf{Macro Strategic Process:} One key component in the MTG is the utility function for each outcome, $u_i^v(z)$, for all $z \in \mathcal{Z}^v$ and $i\in\mathcal{N}$. Utilities represent the payoff or cost of staying or moving to the next node and must be evaluated globally, considering neighboring nodes and their connections. After local exploration and exploitation, the attacker can use obtained credentials or discovered vulnerabilities to move to different nodes, a process known as lateral movement. The attacker's movement and the creation of the attack kill chain depend on the network topology and the expected utilities of each node. We model this decision-making process across the network using an MDP, a Macro Strategic Process (MSP).

The Macro Strategic Process (MSP) for the attacker within the MEGA-PT framework is characterized by a tuple $\Lambda^g = \langle \mathcal{S}, \mathcal{A}^g, T, R, \gamma \rangle$, where $\mathcal{S} = \mathcal{V}$ represents the network nodes as states, $\mathcal{A}^g = \mathcal{E}$ represents the connections between nodes as the action space, $T$ denotes the transition success probability function, $R$ represents the movement rewards, and $\gamma$ is the discounting factor. 
The success of an attack attempt is influenced by the attacker's capability. For the purposes of this analysis, the transition success probability is defined as follows: if the attacker opts to remain at the same node, the probability of staying in the same state is $1$. When the attacker chooses to move to a different node via an outgoing edge, the probability of a successful transition to the new node is denoted by $c_a\in [0,1]$, reflecting the attacker's skill level. If the attack attempt to move is unsuccessful, the attacker remains in the current node. 
The reward function $R$ provides positive rewards $\Bar{V}(v^{\prime})$ based on importance of the node for moving to the next node $v^{\prime} \in\mathcal{V}$ and negative penalties $M_a\in\mathds{R}^-$ for remaining at the same node without progress.

\textbf{Meta-Security Game:} Unlike traditional MDPs, where the attacker can freely choose actions to optimize expected utility, in the realistic penetration testing settings, the attack strategy in the network-level depends on explorations at the local nodes. If the attacker does not find any vulnerabilities leading to the next node, they cannot move forward. Therefore, the global attacks strategy in the MSP relies on the outcomes of the MTG. For each node $v \in \mathcal{V}$, the optimal local penetration plans determine the probability of each tactic outcome, denoted as $\tau^v(z)$, where $z$ represents an outgoing edge from $v$. This probability indicates the likelihood that the attacker will select a particular action $a^g = (v, z)$ in the global attack strategy. For node $v\in\mathcal{V}$, given the MTG $\Gamma^v$ and the local plan profile $\Phi^v$, the global attack strategy is given by
\begin{align}
   \pi^g(a^g = (v,z)\mid s = v) 
    = \tau^v(z\mid \Phi^v),\qquad \forall z\in\mathcal{Z}_v.
    \label{eq:policy}
\end{align}

Policy evaluation estimates the effectiveness of the global attack strategy, $\pi^g$, by calculating expected cumulative utilities. This involves computing value functions using Bellman equations. For each state $s \in \mathcal{S}$, the value function under $\pi^g$ is given by:
{\small
\begin{align}
\label{eq:value}
    V&^{\pi^g}(s) = \\
    &\sum_{a^g \in \mathcal{A}^g} \pi^g(a^g \mid s) \sum_{s' \in \mathcal{V}} T(s' \mid s, a^g) \left[ R(s, a^g, s') + \gamma V^{\pi^g}(s') \right].\notag
\end{align}
}%

This value represents the expected return starting from state $s$ and following the global attack strategy $\pi^g$. In the MTG, the utility of each outcome $z$ at node $v$ reflects the expected reward from taking that action and moving to the next node. The utility functions in the MTG are defined as $u^v_a(z=u) = 
    \sum_{s' \in \mathcal{V}} T(s' \mid s=v, a^g=(v, u)) \left[ R(s, a^g, s') + \gamma V^{\pi^g}(s') \right].$
The defender's utility is the negative of the attacker's utility: $u^v_d(z) = -u^v_a(z)$ for all $z \in \mathcal{Z}^v$.

The exploration at each local MTG generates the global attack strategy, while the estimated value for each node through policy evaluation under the current strategy represent the expected outcome utilities at each MTG. Together, the MSP and the MTGs constitute a meta-security game that captures decision-making in penetration testing at both network and node levels. A detailed example can be found in \Cref{sec:case}.

\begin{definition}[Meta-Security Game]\label{def:metasec}
Given the network system graph $\mathcal{G} = \langle \mathcal{V}, \mathcal{E} \rangle$, the meta-security game is composed of two parts: $\Xi = \langle \{\Gamma^v\}_{v\in\mathcal{V}}, \Lambda^g \rangle$, where $\{\Gamma^v\}_{v\in\mathcal{V}}$ is the set of MTGs and $\Lambda^g$ is the MSP. 
\end{definition}

The MSP and the MTGs are inherently coupled, hence, a holistic solution concept is necessary for the proposed meta-security game.

\begin{definition}[Meta Penetration Playbook]\label{def:playbook}
    Consider the meta-security game $\Xi = \langle \{\Gamma^v\}_{v\in\mathcal{V}}, \Lambda^g \rangle$ defined in Definition~\ref{def:metasec}, the meta penetration playbook $\xi=\langle  \{\Phi^v\}_{v\in\mathcal{V}}, \pi^g \rangle$ is composed of the local penetration profiles at each node and the global attack strategy. They satisfy two conditions: 1) policy Dependency - 
        the global attack strategy $\pi^a$ at the macro strategy process depends on the local penetration plans $\{\Phi^v\}_{v\in\mathcal{V}}$; 2) value Dependency - for each MTG at node $v\in\mathcal{V}$, the utility of each tactic expected outcome depends on the policy evaluation results of global attack strategy $\pi^g$
\end{definition}

\begin{definition}[Network Risk Score]
    Consider the meta-security game $\Xi$ defined in \Cref{def:metasec} and the corresponding meta penetration playbook $\xi$ defined in \Cref{def:playbook}, the network risk score of node $v\in\mathcal{V}$ is defined as $\omega(v\mid \xi)\!\! = \!\!
        \begin{cases}
           \frac{ V^{\pi^g}(v)}{V_{max}} &\! \text{if }V^{\pi^g}(v)\!\geq \!0,\\
           0 &\! \text{otherwise,}
        \end{cases}$
where $V^{\pi^g}(v)$ is the policy evaluation value function under the meta penetration playbook.
\end{definition}

\subsection{Neural-Symbolic Penetration Algorithm}
Consider a partially known meta-security game $\Xi=\langle \{\Gamma^v\}_{v\in\mathcal{V}}, \Lambda^g \rangle$, assuming that unknown information only occurs at the entry point and the edge node of the network. From the attacker's perspective, denote these nodes as the web server $v_o$ and the AI center server $v_d$ that stores and trains ML models. To solve this meta-game with incomplete information, we present the algorithm for solving a meta-penetration testing playbook, followed by an explanation of how neural-symbolic libraries assist in automating adaptation to networks with incomplete information.


\begin{figure}[h]
    \centering   \includegraphics[width=0.75\linewidth]{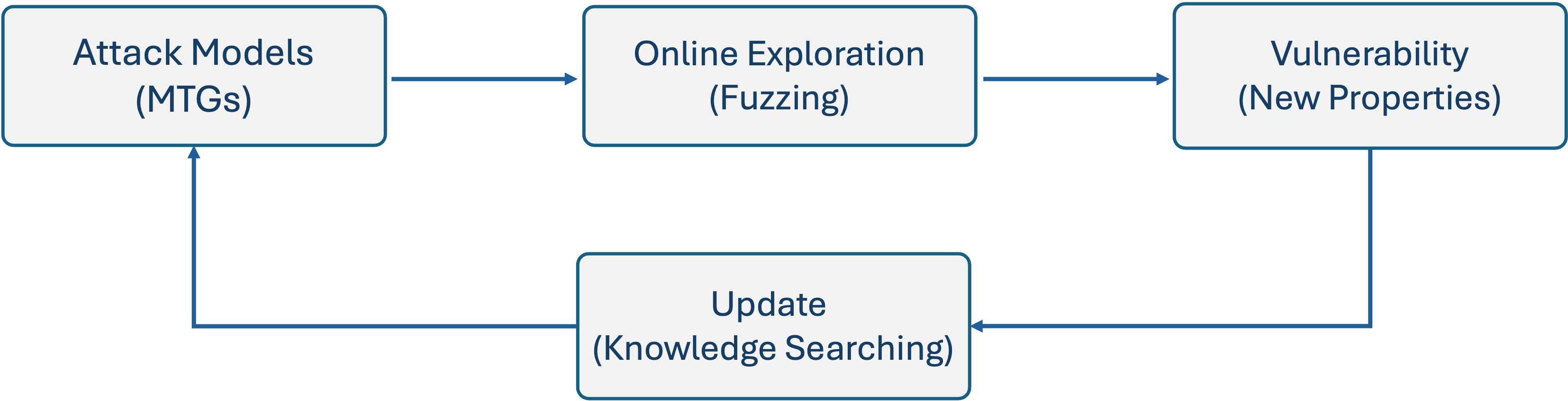}
    \caption{The flow diagram of the symbolic adaptation. Attack models are represented by Multi-Type Graphs (MTGs), and an exploration is performed on this model. When new properties or vulnerabilities in the model are discovered, the model is updated through knowledge searching.}
    \label{fig:NSL}\vspace{-4mm}
\end{figure}

\subsubsection{\textbf{Computation of the Meta Penetration Playbook}}
To describe the solution of our framework, we assume that the meta-security game $\Xi$ operates within a network with complete information. To determine the optimal meta-penetration playbook $\xi$ for $\Xi$ under the chosen solution concept, we propose an iterative computation algorithm in \Cref{alg:compute}. For illustration purposes, we assume that in each MTG, players aim to find the Nash equilibrium as their micro penetration profile $\Phi^{v,eq} = (\sigma^{v,eq}_a, \sigma^{v,eq}_d, \sigma^v_c)$.

\begin{algorithm}
  \caption{Meta Penetration Playbook Computation}\label{alg:compute}
  \begin{algorithmic}[1]
    \State\textbf{Input} Meta-security game $\Xi=\langle \{\Gamma^v\}_{v\in\mathcal{V}},\Lambda^g\rangle$
    \State\textbf{Initialize} Arbitrary global attack strategy $\pi^g$
      \While{Meta penetration playbook not converge}
        \State Obtain the value function $V^{\pi^g}$ of $\Lambda^g$ by \eqref{eq:value}
        \State Update the $u^v_i$ in each MTG $\Gamma^v$
        \State For all $v\in\mathcal{V}$ compute $\Phi^{v,eq} =(\sigma^{v,eq}_a,\sigma^{v,eq}_d,\sigma^v_c)$\
        \State Update $\pi^a$ under $\{\Phi^{v,eq}\}_{v\in\mathcal{V}}$ using \eqref{eq:policy}
      \EndWhile
      \State \textbf{Return} Meta penetration playbook $\xi=\langle  \{\Phi^{v,eq}\}_{v\in\mathcal{V}},$ $ \pi^g \rangle$
  \end{algorithmic}
\end{algorithm}
For the extensive form game tree design, we can use backward induction to find the subgame perfect Nash equilibrium. Alternatively, a neural network-based approach can be employed to find the game solution if the MTG is computationally complicated.

\subsubsection{\textbf{Symbolic Adaptation of MTGs}}
In the meta-security game with incomplete information, our goal is to explore the MTGs with partially or completely unknown information, and the flow diagram is shown in \Cref{fig:NSL}. As an example, we consider the MTGs of the entry node and the edge node in the network, denoted as $\Gamma^{v_o}$ and $\Gamma^{v_d}$, respectively. We use First-Order-Logic (FOL) representation $\kappa^v$ \cite{LeBouthillier1999SymbolicAI} to describe the properties of the network servers in \Cref{tab:FOL}.

{\renewcommand{\arraystretch}{1}
\begin{table}[!htbp]
\centering
\footnotesize
\begin{tabular}{|l|l|}
\hline
\textbf{MTG Properties} & \textbf{FOL Representation} \\
\hline
Windows OS & $\text{is-windows}(v)$ \\
Microsoft protected & $\text{is-microsoftD}(v)$ \\
Allow Authentication Bypass & $\text{can-bypassD}(v)$ \\
$\cdots$ & $\cdots$ \\
\hline
\end{tabular}
\caption{Node/Device Properties and Their Corresponding FOL Representations}\label{tab:FOL}\vspace{-2mm}
\end{table}}

For each node $v\in\mathcal{V}$, we assume that there is a finite property set that captures all possible configurations on the node, denoted by $\mathcal{K}^v:=\{\kappa^v_l\}_{l\in[1,\cdots|\mathcal{K}^v|]}$.  Property exploration refers to the expansion and updating of the property set $\mathcal{K}^v$ associated with node $v$. In the web server, assuming it is using a Linux operating system and has a Microsoft Defender, the properties are represented as $\mathcal{K}^{v_o}=\{\text{is-linux}(\cdot), \text{is-microsoftD}(\cdot)\}$. Assume the penetration tester is using an automated fuzzing technique \cite{Gorbunov2010AutoFuzzAN} to explore the node server and discovers a new vulnerability that allows attacker bypass authentication check. Then, the updated property set of the node is $\mathcal{K}^{v_o,\prime}=\{\text{is-linux}(\cdot), \text{is-microsoftD}(\cdot), \text{can-bypassD}(\cdot)\}$. Following the same logic, we can describe an AI server using the Windows system with the Windows Defender's Protection, where an ethical attacker intends to perform an evasion attack on the ML model tactic technique (detail will be described in the following section), as $\mathcal{K}^{v_d}=\{\text{is-windows}(\cdot), \text{is-windowsD}(\cdot), \text{evadeMLmodel}(\cdot)\}$.

We use a knowledge library to store all MTGs related to node $v\in\mathcal{V}$, formally defined as follows. 
\begin{definition}[Knowledge Library]\label{def:Knowledge}
Consider that a knowledge library $\mathcal{L}^v$ is a set of MTGs corresponding to a specific node $v$, defined as $\mathcal{L}^v := \{\Gamma^v(\mathcal{K}^v)\}_{\mathcal{K}^v \in POW(\mathcal{P}^v)}$, where $\mathcal{P}^v$ contains all possible node properties for $v$ and $POW(\mathcal{P}^v) =\{U|U\subseteq \mathcal{P}^v\}$ is the power set of $\mathcal{P}^v$.
\end{definition}
For given property set $\mathcal{K}^v$ and knowledge library $\mathcal{L}^v$, a symbolic adaptation process is to find a $\Gamma^v_i\in \mathcal{L}^v$ that satisfies:
{\small
\begin{align}
\left(\bigwedge_{l=1}^{|\mathcal{K}^v|} \kappa^v_l(\Gamma^v_i)\right) \wedge \left(\bigwedge_{\kappa^v \in \mathcal{P}^v \setminus \mathcal{K}^v} \neg\kappa^v(\Gamma^v_i)\right) \text{ is true.}\label{eq:sk_adapt}
\end{align}
}
Here, we assume that for any possible property set $\mathcal{K}^v$, there exist a feasible node description of $v$ that satisfies \eqref{eq:sk_adapt}. By defining the solution of the penetration playbook as neural and the adaptation of node properties as symbolic, we present a conceptual workflow of ADAPT in \Cref{alg:ADAPT}.

\begin{algorithm}
  \caption{Neural-Symbolic Penetration Library}\label{alg:ADAPT}
  \begin{algorithmic}[1]
    \State \textbf{Input:} MSP $\Lambda^{g}$, global graph $\mathcal{G}=\langle \mathcal{V}, \mathcal{E} \rangle$, knowledge libraries $\{\mathcal{L}^v\}_{v\in\mathcal{V}}$, and the initial property set of nodes $\{\mathcal{K}^v\}_{v\in\mathcal{V}}$
    \State \textbf{Symbolic Security Game Adaptation:}
    \For{$v\in\mathcal{V}$}
        \While{$\mathcal{K}^v$ is still evolving}
        \State Explore node $v$, find a FOL property $\kappa^v_{new}$
        \If{$\kappa^v_{new}\notin\mathcal{K}^v$}
        \State $\mathcal{K}^v=\mathcal{K}^v\cup \kappa^v_{new}$
        \EndIf
        \EndWhile
    \State Obtain $\Gamma^v$ by $\mathcal{K}^v$ with \eqref{eq:sk_adapt}
    \EndFor
    \State Construct meta-security game $\Xi = \langle \{\Gamma^v\}_{v\in\mathcal{V}}, \Lambda^g \rangle$
    \State \textbf{Neural Penetration Playbook Computation:}
    \State Obtain the meta-penetration playbook $\xi$ using \Cref{alg:compute}
    \State \textbf{Return} the explored meta-security game $\Xi$, and the meta penetration playbook $\xi$.
  \end{algorithmic}
\end{algorithm}
Since the knowledge library may be imperfect and unable to fully describe the node properties, 
there is a need to constantly update the knowledge library. Due to the limited space, we leave the discussion for future work.

\begin{figure}[!htbp]\vspace{-4mm}
    \centering   \includegraphics[width=0.75\linewidth]{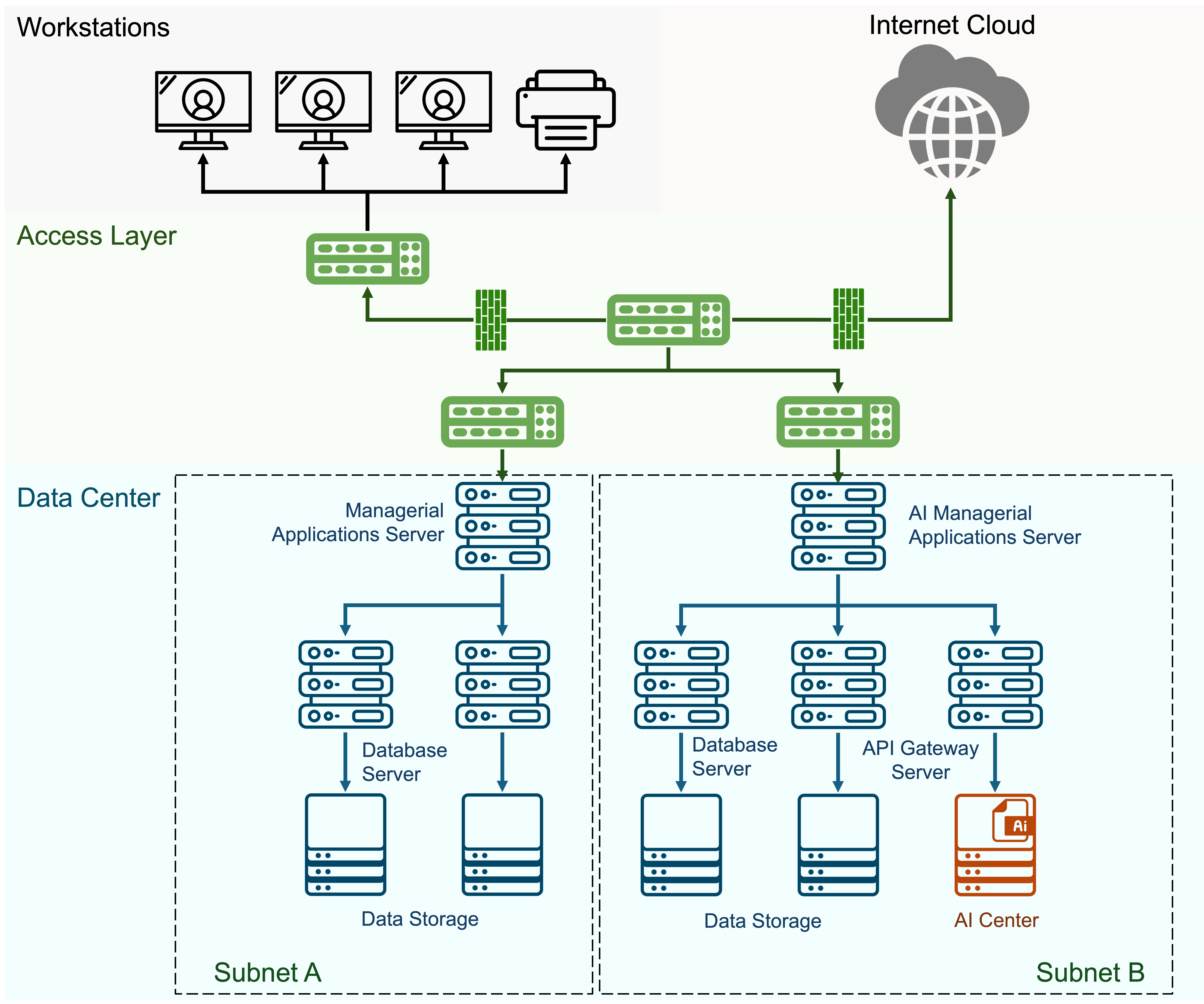}
    \caption{An illustration of a hospital network with AI data center, consisting of four different areas (Internet Cloud, workstations, access layer, and data center) that represent the transition from the public network to the protected private network of the hospital. The data center includes two different subnets. One is for only data storage, and the other is for both data storage and AI applications. The attacker will access the network from the web server in workstations and target the AI center as the network's critical asset.}
    \label{fig:hospital_network}\vspace{-4mm}
\end{figure}

\section{Hospital Case Study}\label{sec:case}
In this section, we discuss a case study of a typical IT infrastructure within a healthcare environment, i.e., a hospital network architecture shown in \Cref{fig:hospital_network}, referring to \cite{Salleh2014NetworkArchitecture} and \cite{OnyxUSA2024HardwareArchitecture}. The attacker accesses the hospital network from a web server and targets the AI center in subnet B. Through this example, we aim to highlight the uniqueness of the MTG design in an AI-enhanced network and emphasize the guidance and value of our algorithm for users in such networks.

Real-life examples show the ongoing need for hospitals to deploy AI servers locally \cite{Krass2023ProfessionalAI}, typically on the edge device. Consider a hospital deploying an AI model for disease diagnosis using an ML technique similar to those mentioned in the review paper \cite{Ahsan2022MachineLearningDiagnosis}. We assume a malicious attacker (entering the network from the web server node in workstations) aims to steal the access authority of the edge node AI center and perform their attack within the network described in \Cref{fig:hospital_network}. For simplification, the topology of the Meta-security game for this network is considered fixed, except for the edge node AI center, which contains the model development and deployment procedures. The MTG in the edge node AI center is determined by the different impact techniques.

\subsection{Penetration Path to access the AI Center}
\begin{figure}[!htbp]
    \vspace{-4mm}\centering   \includegraphics[width=0.8\linewidth]{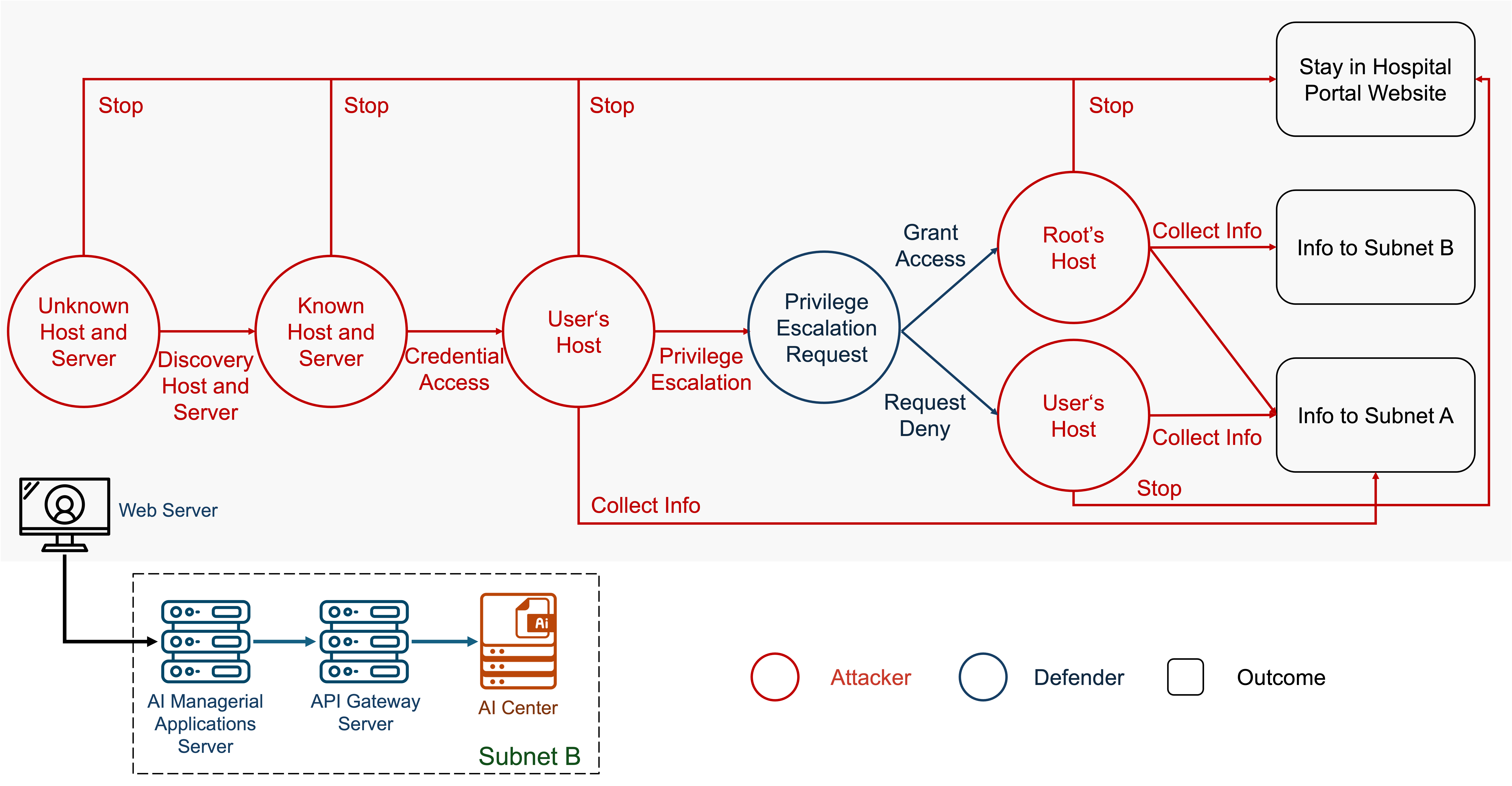}
    \caption{An illustration of the meta-security game for in the hospital network. For simplification, only the path to the critical asset, i.e., the AI center, is considered. The local penetration plans in the micro games influence the global attack strategy, while the policy evaluation in the macro process helps provide the utilities for the micro games.}
    \label{fig:access_ai}\vspace{-4mm}
\end{figure}

\Cref{fig:access_ai} illustrates an example of the meta-security game in the hospital network shown in \Cref{fig:hospital_network}. The attacker starts from the web server and tries to access the AI center by penetrating the AI managerial applications server and the API gateway server. To conserve space in this paper, we present the MTG of the web server following the MITRE ATT\&CK framework \cite{mitre2020mitigations} as an example and omit the others. The MSP defines the global attack strategy by forming an attack kill chain and providing estimated values for each node through policy evaluation under the current strategy. These estimated values represent the expected outcome utilities at each MTG, guiding the formulation of detailed penetration plans at each node. The sequence of attack and defense techniques at the local node influences the global attack strategy from a macro perspective, emphasizing how lateral movement is determined by exploration and exploitation. By continuing the iteration of this process, a meta-solution, i.e., an optimal penetration playbook, can be reached.

\subsection{Impact Techniques}

The goal for the attacker is to interrupt, erode confidence in, or destroy the machine learning systems and data. Traditionally, Techniques used for impact can include destroying or tampering with data. Following the MITRE ATLAS framework \cite{ATLAS2024AdversarialThreat}, we use examples to illustrate the differences between three different impact techniques' MTG trees.

\textbf{Erode ML Model Integrity:} Attacks can happen during the model training, i.e., damage on the model integrity. To influence the ML model's integrity, there are two primary approaches. The first approach is to directly alter the model, which is efficient and straightforward but also more likely to be detected \cite{Lakshmanan2024SleepyPickle}. The second approach is to poison the training data \cite{YERLIKAYA2022118101},\cite{sample_attack}, thereby degrading the model's performance—often summarized as "garbage in, garbage out."

\textbf{Evade ML Model:} Attacks can also occur during model access. An attacker can craft adversarial data that prevents a machine learning model from correctly identifying the contents of the data. This technique can be used to evade downstream tasks where machine learning is utilized. One example of this type of impact technique is the bypassing of Cylance antivirus products \cite{Cylance2019IKillYou}. As a result, this impact can cause the adversary's desired effect on the target model, leading to consequences such as misclassification, missed detections, or maximized energy consumption.

\begin{figure}[!htbp]
    \centering   \includegraphics[width=0.75\linewidth]{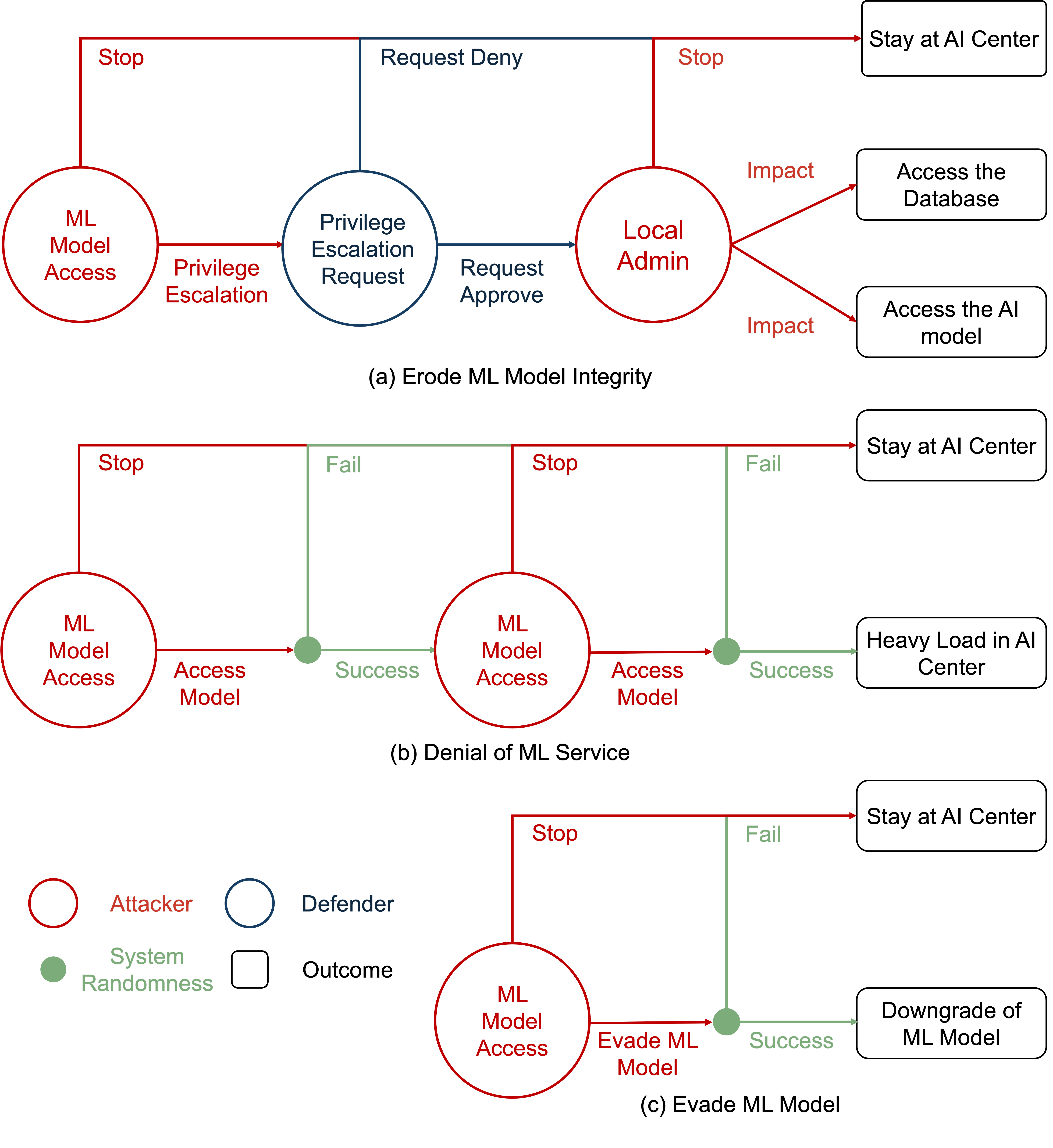}
    \caption{An illustration of MTGs for different impact techniques. Red nodes represent the attacker, blue nodes represent the defender, yellow nodes represent the system randomness, and green nodes represent outcomes.}
    \label{fig:exp_impact}\vspace{-4mm}
\end{figure}

\textbf{Denial of ML Service:} By this impact technique, the attacker is targeting ML systems' accessibility instead of integrity. The attacker focuses on generating a flood of requests for degrading or shutting down the service. Recall the massive disaster that occurred in July 2024, when the “blue screen of death” affected most Microsoft computers in airports and hospitals \cite{NYT2024ChaosConfusion}. Although this incident was caused by a system bug in Microsoft software, its impact was similar to that of a denial-of-ML-service attack. \Cref{fig:exp_impact} shows MTG trees for three different impact techniques. \vspace{-4mm}


\subsection{Numerical Experiments}
\begin{figure}[!htbp]
    \centering   \includegraphics[width=0.9\linewidth]{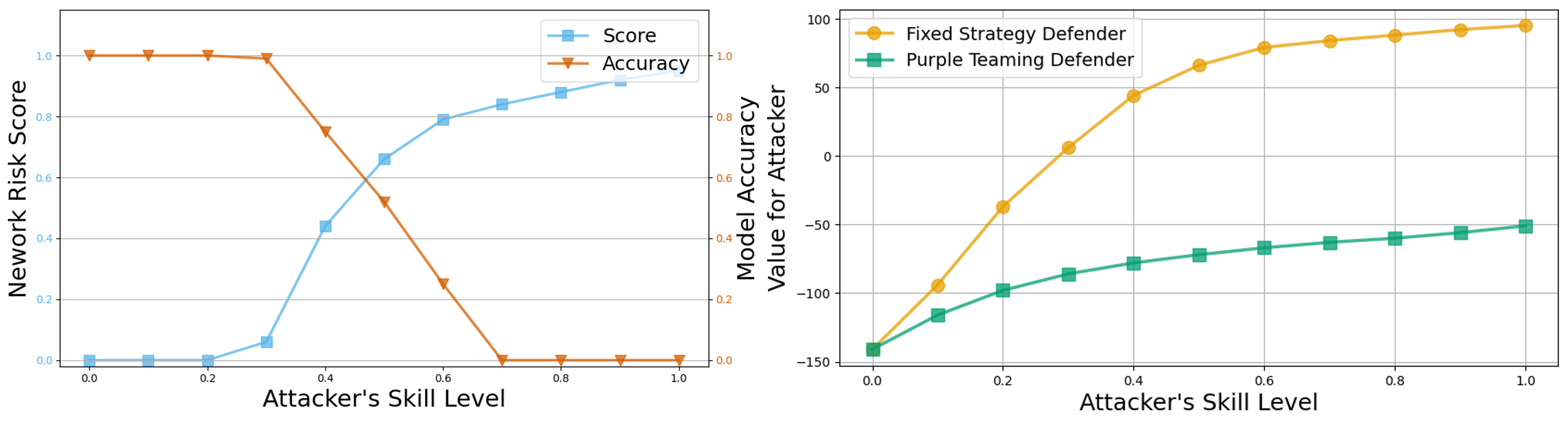}
    \caption{Left: Experimental results for the network risk score and ML model accuracy under different attacker's skill levels with a fixed strategy defender. The blue line represents the network risk score, while the red line indicates the ML model accuracy. Right: The optimal attacker's value of the entry node under different attacker's skill levels. The orange line represents the fixed strategy defender, while the green line represents the purple teaming defender.}\vspace{-4mm}
    \label{fig:exp}
\end{figure}

This experiments section provides insight into the risk assessment process of penetration testing in an AI-enhanced hospital network. Assume the attacker chooses to erode the integrity of a regression ML model (using a support vector machine in this case) through a data poisoning algorithm as described in \cite{YERLIKAYA2022118101}, which can degrade the model's accuracy. We maintain the hospital network configuration as shown in \Cref{fig:hospital_network} and assume this ML model is placed in the AI center node of subnet B as the critical asset of the network. Let the risk score $\omega(v\mid \xi)$ of the entry node be directly identical to the data poisoning ratio, and denote the network uses non-sophisticated defenders that only use fixed strategies. By iterating the attacker's skill level from $0$ to $1$ with an interval of $0.1$, we obtain the experimental results shown in the left of \Cref{fig:exp}.\vspace{-4mm}

We show all possible risk scores and model accuracy at different levels of the attacker's skill. As the attacker's skill level increases, the risk score rises, leading to a corresponding decrease in the ML model's accuracy. For a $c_a=1$ attacker, they can obtain nearly $100\%$ of the value from the critical asset and cause complete damage to the ML model.

The experiment provides a risk evaluation that helps users determine whether they need to increase their investment in preparation for future attacks and illustrates the impact of reachability on hospital security. For instance, refer to the left of \Cref{fig:exp}, if a user can tolerate a maximum accuracy loss of around $20\%$ for the ML model, they will likely increase their defense budget if they believe that future attackers will have a high skill level, specifically when $c_a \geq 0.4$, vice versa. This demonstrates the vulnerability of a fixed-strategy defender when confronted with a strategic attacker. Let us consider a more strategic defender, such as a purple teaming defense, which is a collaborative cybersecurity strategy that integrates both offensive (red teaming) and defensive (blue teaming) perspectives to strengthen the overall security posture of the system. Suppose we aim to use a Stackelberg equilibrium to find an optimal purple teaming meta-penetration playbook. Let $\sigma^{v,*}_a$ be the anticipated optimal penetration plan for the attacker given the defense plan (it can be a fixed defense strategy or the defender's current strategy). The optimal purple teaming defense plan $\sigma^{v,pur}_d$ is given by $\sigma^{v,pur}_d(\sigma^v_c) \in \max_{\sigma^v_d \in \Sigma^v_d}\quad  u^v_d(\sigma^{v,*}_a, \sigma^v_d,\sigma^v_c)$ subject to $\sigma^{v,*}_a \in \arg\max_{\sigma^v_a\in\Sigma^v_a}u^v_a(\sigma^v_a, \sigma^v_d,\sigma^v_c)$.

It is evident that the purple teaming defense is more sophisticated than a fixed strategy. The experimental results shown in the right of \Cref{fig:exp} demonstrate the effectiveness of the purple teaming defense with $V_{max}=100$ and a penalty of $-50$ for staying in a node. Regardless of the attacker's skill level, their values always remain below $0$, indicating that the defender has the advantage. As a trade-off, the purple teaming defender requires more computing resources. In the hospital context, a purple teaming defender can be seen as an upgrade from a fixed strategy defender, meaning the hospital has invested more budget in cybersecurity defense.






\vspace{-1mm}
\section{Conclusion}
In this work, we propose ADAPT, a game-theoretic and neuro-symbolic framework for automated distributed adaptive penetration testing. ADAPT addresses the critical demands of modern healthcare infrastructure and has the ability to dynamically adjust to evolving threats, and its effectiveness in securing AI systems against zero-day vulnerabilities underscores its potential as a vital tool for enhancing cybersecurity in healthcare environments. The case study demonstrates a real-life hospital network structure and showcases the framework’s effectiveness in addressing various tactical techniques employed by attackers. In numerical experiments, the results show around 98\% improvement in reducing the risk score of the system by using the purple teaming defender as the replacement of a fixed strategy defender.


\bibliographystyle{IEEEtran}
\bibliography{IEEEabrv,mybibfile}

\end{document}